\documentclass[twocolumn,showpacs,preprintnumbers,amsmath,amssymb]{revtex4}

\usepackage{graphicx}
\usepackage{dcolumn}
\usepackage{bm}

\begin{document}

\preprint{}

\title{Spin Hall effect of light in photon tunneling}
\author{Hailu Luo}
\author{Shuangchun Wen}\email{scwen@hnu.cn}
\author{Weixing Shu}
\author{Dianyuan Fan}
\affiliation{Key Laboratory for Micro/Nano Opto-Electronic Devices
of Ministry of Education, School of Information Science and
Engineering, Hunan University, Changsha 410082, People's Republic of
China}
\date{\today}

\begin{abstract}
We resolve the breakdown of angular momentum conservation on
two-dimensional photon tunneling by considering spin Hall effect
(SHE) of light. This effect manifests itself as
polarization-dependent transverse shifts of field centroid when a
classic wave packet tunnels through a prism-air-prism barrier. For
the left or the right circularly polarized component, the transverse
shift can be modulated by altering the refractive index gradient
associated with the two prisms. We find that the SHE in conventional
beam refraction can be evidently enhanced via photon tunneling
mechanism. The transverse spatial shift is governed by the total
angular momentum conservation law, while the transverse angular
shift is governed by the total linear momentum conservation law.
These findings open the possibility for developing new nano-photonic
devices and can be extrapolated to other physical systems.
\end{abstract}

\pacs{42.25.-p, 42.79.-e, 41.20.Jb}
\keywords{spin Hall effect of light, photon tunneling, angular
momentum}

\maketitle

\section{Introduction}\label{SecI}
The tunneling effect is a cornerstone of both quantum mechanics and
classical electrodynamics, and presents intriguing features that
stimulate an ongoing interest in nano-photonics~\cite{Novotny2006}.
From the viewpoint of quantum mechanics, wave or particle may
penetrate through a classically impenetrable barrier. In particular,
frustrated total internal reflection (FTIR) is considered as a
classical analogy of quantum-mechanical tunneling. In this case,
optical fields penetrate across an air gap between two adjacent
glass prisms, giving transmission of light beyond the critical
angle~\cite{Chiao1991,Steinberg1994}. Until now, it is generally
believed that the photon tunneling is a two-dimensional
process~\cite{Steinberg1993,Balcou1997,Carey2000,Winful2003,Hooper2006}:
Tunneling only occurs in the incidence plane. However, the total
angular momentum is unconserved in the two-dimensional
FTIR~\cite{Player1987,Fedoseyev1988,Fedoseyev2009}. This difficulty
arises from the neglect of orbital angular momentum caused by
transverse spatial shift of field
centroid~\cite{Fedorov1965,Imbert1972}. Therefore, there is a
apparent breakdown of angular momentum conservation in
two-dimensional photon tunneling. In our opinion, this discrepancy
may be resolved by considering spin Hall effect (SHE) of light.

The SHE of light is a photonic version of SHE in electronic
systems~\cite{Murakami2003,Sinova2004,Wunderlich2005}, in which the
spin of photons play the role of the spin of charges, and a
refractive index gradient acts as the electric potential
gradient~\cite{Onoda2004,Bliokh2006a,Hosten2008}. Such a spatial
gradient for the refractive index could occur at an interface
between two materials. Recently, the SHE of light has also been
observed in glass cylinder with gradient refractive
index~\cite{Bliokh2008}, in scattering from dielectric
spheres~\cite{Haefner2009}, on the direction tilted with respect to
beam propagation axis~\cite{Aiello2009}, and in silicon via
free-carrier absorption~\cite{Menard2010}. This interesting effect
manifests itself as the split of a linearly polarized beam into two
left circularly and right circularly components. The splitting in
the SHE, implied by angular momentum conservation, takes place as a
result of an effective spin-orbit interaction. In a symmetric
prism-air-prism barrier, we predict that the polarization-dependent
transverse shifts are very small. This is a possible reason why the
tiny scale of the effect escaped detection in photon tunneling
experiments.

In this work, we use an air gap between two glass prisms as the
potential barrier to explore the SHE in photon tunneling. First,
starting from the representation of a plane-wave angular spectrum,
we establish a general propagation model to describe the photon
tunneling. Based on this model, the SHE of light can be obtained
very simply by calculating the transverse shifts. Next, we explore
what happens to the SHE in photon tunneling and find that the SHE
can be evidently enhanced with an asymmetric barrier structure.
Then, we examine what roles the refractive index gradient of the two
prisms plays in the SHE of light. Our result shows that this
interesting effect can be modulated by altering the refractive index
gradient. Finally, we attempt to reveal the inherent secret
underlying the SHE in photon tunneling. The polarization-dependent
transverse shifts governed by the total momentum conservation law,
provides us a new way to clarify the nature of photon tunneling
effect in three dimensions. We believe that the study of the SHE of
light may provide insights into the fundamental properties of photon
tunneling.

\section{Photon tunneling model}\label{SecII}
\begin{figure}
\includegraphics[width=8cm]{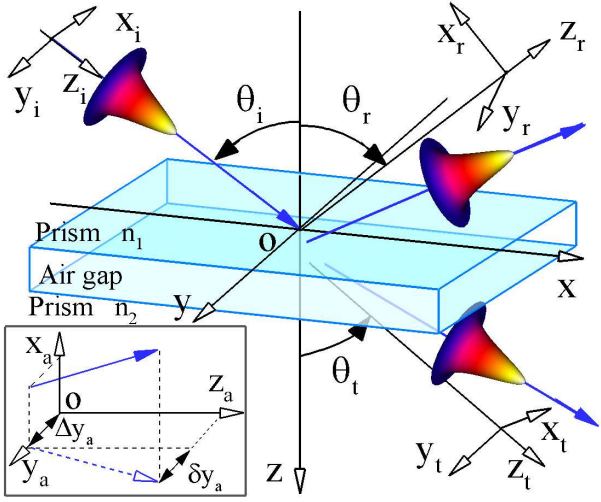}
\caption{\label{Fig1} (color online) Three-dimensional geometry of
beam reflection and tunneling from a prism-air-prism barrier. The
inset shows that the $a$th beam undergoes a transverse spatial shift
$\Delta y_a$ and a transverse angular shift $\delta y_a$.}
\end{figure}

We begin by establishing a three-dimensional propagation model to
describe photon tunnels through a prism-air-prism barrier.
Figure~\ref{Fig1} illustrates the scheme of our optical tunneling.
The $z$ axis of the laboratory Cartesian frame ($x,y,z$) is normal
to the prism-air interface locating at $z=0$. We use the coordinate
frames ($x_a,y_a,z_a$) for individual beams, where $a=i,r,t$ denotes
incident, reflected, and tunneling beams, respectively. A beam
impinges from the first prism onto an air gap with a thickness of
$d$. For incidence angles $\theta_i$ greater than the critical angle
of total internal reflection: $\theta_{ic}=\sin^{-1}(1/n_1)$, most
of photons are reflected, and part of them tunnel through the air
gap.

In general, the electric field of the $a$th beam can be solved by
employing the Fourier transformations~\cite{Goodman1996}. The
complex amplitude can be conveniently expressed as
\begin{eqnarray}
\mathbf{E}_a(x_a,y_a,z_a )&=&\int d k_{ax}dk_{ay}
\tilde{\mathbf{E}_a}(k_{ax},k_{ay})\nonumber\\
&&\times\exp [i(k_{ax}x_a+k_{ay}y_a+ k_{az} z_a)],\label{noapr}
\end{eqnarray}
where $k_{az}=\sqrt{k_a^2-(k_{ax}^2+k_{ay}^2)}$. The approximate
paraxial expression for the field in Eq.~(\ref{noapr}) can be
obtained from the expansion of the square root of $k_{az}$ to the
first order~\cite{Lax1975}, which yields
\begin{eqnarray}
\mathbf{E}_a&=&\exp(i k_a z_a) \int dk_{ax}dk_{ay}\tilde{\mathbf{E}_a}(k_{ax},k_{ay})\nonumber\\
&&\times\exp \left[i\left(k_{ax}x_a+k_{ay}y_a-\frac{k_{ax}^2+k_{a
y}^2}{2 k_a}z_a\right)\right].\label{apr}
\end{eqnarray}
In the first prism, we consider an arbitrarily polarized Gaussian
beam propagating parallel to the positive $z_i$ axis. The angular
spectrum of electric field amplitude of such a beam can be written
as
\begin{equation}
\tilde{\mathbf{E}}_i=(\alpha\mathbf{e}_{ix}
+\beta\mathbf{e}_{iy})\exp\left[-\frac{z_R(k_{ix}^2+k_{iy}^2)}{2
k_0}\right]\label{asi}.
\end{equation}
Here, $z_{R}=k_0 w_0^2 /2$ is the Rayleigh length in free space and
$k_0=\omega/c$ is the wave number. The coefficients $\alpha$ and
$\beta$ satisfy the relation
$\sigma_i=i(\alpha\beta^\ast-\alpha^\ast\beta)$. The polarization
operator $\sigma_i=\pm1$ corresponds to left- and right-handed
circularly polarized light, respectively. It is well known that
circularly polarized Gaussian beam can carry spin angular momentum
$\pm1\hbar$ per photon due to its polarization
state~\cite{Beth1936}. Substituting Eq.~(\ref{asi}) into
Eq.~(\ref{apr}) provides the expression of the incident field:
\begin{eqnarray}
\mathbf{E}_i(x_i,y_i,z_i)&\propto&(\alpha\mathbf{e}_{ix}+\beta\mathbf{e}_{iy})
\exp(ik_{iz}z_i)\nonumber\\
&&\times\exp\left[-\frac{n_1 k_0}{2}\frac{x_i^2+y_i^2}{n_1  z_{R}+ i
z_i}\right]\label{fieldi}.
\end{eqnarray}
The calculation of the reflected and tunneling fields requires the
explicit solution of the boundary conditions at the interfaces.
Thus, we need to know the generalized Fresnel reflection and
transmission coefficients of the barrier which read as
\begin{equation}
r_{A}=\frac{R_{A}+R'_{A}\exp(2ik_0\sqrt{1-n_1^2 \sin^2\theta_i}
d)}{1+R_{A}R'_{A}\exp(2ik_0\sqrt{1-n_1^2 \sin^2\theta_i}
d)}\label{ra},
\end{equation}
\begin{equation}
t_{A}=\frac{T_{A}T'_{A}\exp(i k_0\sqrt{1-n_1^2 \sin^2\theta_i}
d)}{1+R_{A} R'_{A}\exp(2ik_0\sqrt{1-n_1^2 \sin^2\theta_i}
d)}\label{ta}.
\end{equation}
Here, $A\in\{p, s\}$, $R_{A}$ and $T_{A}$ are the Fresnel reflection
and transmission coefficients at the first interface, respectively.
$R'_{A}$ and $T'_{A}$ are the corresponding coefficients at the
second interface.

We first explore the reflected field in the first prism. The
reflected angular spectrum $\tilde{E}_r(k_{rx},k_{ry})$ is related
to the boundary distribution of the electric field by means of the
relation:
\begin{eqnarray}
\tilde{\mathbf{E}}_r=\left[
\begin{array}{cc}
r_p &\frac{k_{ry} \cot\theta_i}{k_0} (r_p+r_s) \\
-\frac{k_{ry} \cot\theta_i}{k_0} (r_p+r_s) & r_s
\end{array}
\right]\tilde{\mathbf{E}}_i.\label{matrixr}
\end{eqnarray}
From the Snell's law, we can get $k_{rx}=-k_{ix}$ and $k_{ry}=
k_{iy}$. In fact, after the angular spectrum on the plane $z_r=0$ is
known, Eq.~(\ref{apr}) together with Eqs.~(\ref{asi}) and
(\ref{matrixr}) provides the expression of the field on the plane
$z_r>0$ as
\begin{eqnarray}
\mathbf{E}_{r}&\propto&\bigg[\alpha r_p\left(1-i\frac{x_r}{n_1 z_R+
i z_r}\frac{\partial \ln r_p}{\partial
\theta_i}\right)\nonumber\\
&&+i\beta(r_p+r_s)\frac{y_r}{n_1 z_R+ i z_r}\cot\theta_i \bigg]\nonumber\\
&&\times\exp\left[-\frac{n_1 k_0}{2}\frac{x_r^2+y_r^2}{n_1 z_R+
i z_r}\right]\mathbf{e}_{rx}\nonumber\\
&&+\bigg[\beta r_s\left(1-i\frac{x_r}{n_1 z_R+ i z_r}\frac{\partial
\ln r_s}{\partial \theta_i}\right)\nonumber\\
&&-i\alpha(r_p+r_s)\frac{y_r}{n_1 z_R+ i z_r}\cot\theta_i
\bigg]\nonumber\\
&&\times\exp\left[-\frac{n_1 k_0}{2}\frac{x_r^2+y_r^2}{n_1 z_R+ i
z_r}\right]\mathbf{e}_{ry}.\label{fieldr}
\end{eqnarray}
For the reflection on prism-air-prism barrier, the spatial profile
of the reflected beam experiences an alteration.

We next explore the transmitted or tunneling field in the second
prism. Whether the field in the second prism is transmitted or
tunneling depends on the incident angle: In the transmission case,
$\theta_i<\theta_{ic}$; while in the tunneling case,
$\theta_i>\theta_{ic}$. The tunneling field can be derived which
does not require knowledge of the fields inside the air
gap~\cite{Novotny2006}. The transmitted angular spectrum
$\tilde{E}_t(k_{tx},k_{ty})$ related to the boundary distribution of
the electric field is written as:
\begin{eqnarray}
\tilde{\mathbf{E}}_t=\left[
\begin{array}{cc}
t_p &\frac{k_{ty} \cot\theta_i}{k_0} (t_p-\eta t_s) \\
\frac{k_{ty} \cot\theta_i}{k_0} (\eta t_p-t_s)& t_s
\end{array}
\right]\tilde{\mathbf{E}}_i,\label{matrixt}
\end{eqnarray}
where $\eta=\cos\theta_t/\cos\theta_i$. From the Snell's law under
the paraxial approximation, we have $k_{tx}=k_{ix}/\eta $ and
$k_{ty}= k_{iy}$. Substituting Eqs.~(\ref{asi}) and (\ref{matrixt})
into Eq.~(\ref{apr}), the transmitted or tunneling field is given by
\begin{eqnarray}
\mathbf{E}_{t}&\propto&\exp\left[-\frac{n_2
k_0}{2}\left(\frac{x_t^2}{z_{Rx}+ i z_t}+\frac{y_t^2}{z_{Ry}+ i
z_t}\right)\right]\nonumber \\&&\times\bigg[\alpha
t_p\left(1+i\frac{n_2 \eta x_t}{z_{Rx}+ i z_t}\frac{\partial \ln
t_p}{\partial\theta_i}\right)\nonumber\\
&&+i\beta\frac{n_2 y_t}{z_{Ry}+ i z_t}(t_p-\eta t_s)\cot\theta_i \bigg]\mathbf{e}_{tx}\nonumber\\
&&+\exp\left[-\frac{n_2 k_0}{2}\left(\frac{x_t^2}{z_{Rx}+ i
z_t}+\frac{y_t^2}{z_{Ry}+ i
z_t}\right)\right]\nonumber\\
&&\times\bigg[\beta t_s\left(1+i\frac{n_2 \eta x_t}{z_{Rx}+ i
z_t}\frac{\partial \ln
t_s}{\partial\theta_i}\right)\nonumber\\
&&+i\alpha \frac{n_2 y_t}{z_{Ry}+ i z_t}(\eta t_p-t_s)\cot\theta_i
\bigg]\mathbf{e}_{ty}. \label{fieldt}
\end{eqnarray}
An interesting point to be noted is that there are two different
Rayleigh lengths: $z_{Rx}=n_2\eta^2k_0w_0^2/2$ and $z_{Ry}=n_2 k_0
w_0^2/2$, characterizing the spreading of the beam in the direction
of $x_t$ and $y_t$ axes, respectively. Up to now, we have
established a general propagation model to describe the SHE of
light in three-dimensional photon tunneling. This intriguing effect
manifests itself as polarization-dependent transverse shifts of beam
centroid. As a result, the photon tunneling is no longer a two
dimensional process.

\section{Spin Hall effect of light}\label{SecIII}
To reveal the SHE of light, we now determine the
polarization-dependent transverse shifts of beam centroid. The
intensity distribution of electromagnetic fields is closely linked
to the Poynting vector~\cite{Jackson1999}
$I(x_a,y_a,z_a)\propto\mathbf{S}_a\cdot \mathbf{e}_{az}$. Here, the
Poynting vector is given by
$\mathbf{S}_a\propto\text{Re}[\mathbf{E}_a^\ast\times\mathbf{H}_a]$
and the magnetic field can be obtained by $\mathbf{H}_a=-i
k^{-1}\nabla\times\mathbf{E}_a$. At any given plane
$z_a=\text{const.}$, the transverse shifts of beam centroid are
given by
\begin{equation}
\langle y_{a}\rangle= \frac{\int \int y_a I(x_a,y_a,z_a) \text{d}x_a
\text{d}y_a}{\int \int I(x_a,y_a,z_a) \text{d}x_a
\text{d}y_a}.\label{centroid}
\end{equation}
Here, $\langle y_{a}\rangle$ can be written as a combination of
$z_a$-independent term and $z_a$-dependent term: $\langle
y_{a}\rangle=\Delta y_{a}+\delta y_{a}$. The former gives transverse
spatial shifts while the latter can be regarded as transverse angular
shifts.

We first consider the transverse shift of the incident field.
Substituting Eq.~(\ref{fieldi}) into Eq.~(\ref{centroid}), we get
\begin{equation}
\Delta y_i=0,~~~~~\delta y_i=0\label{TLSI}.
\end{equation}
This simple result shows that there does not exist a transverse
spatial shift or transverse angular shift in a fundamental Gaussian
beam~\cite{Luo2010}. We next consider the transverse shift of the
reflected field. Substituting Eq.~(\ref{fieldr}) into
Eq.~(\ref{centroid}), we have
\begin{eqnarray}
\Delta y_r &=&-\frac{1}{k_0}\frac{f_p f_s \cot \theta_i }{|r_p|^2
f_p^2+|r_s|^2 f_s^2}\Bigl[\bigl(|r_p|^2 + |r_s|^2\bigr)\sin\psi\nonumber\\
&&+ 2 |r_p||r_s|\sin(\psi-\phi_p+\phi_s )\Bigr]\label{TSR},
\end{eqnarray}
\begin{equation}
\delta y_r=\frac{z_r}{k_0 z_R}\frac{f_p f_s  (|r_p|^2 -
|r_s|^2)\cot\theta_i \cos\psi}{|r_p|^2 f_p^2+|r_s|^2
f_s^2}\label{TLSR}.
\end{equation}
Here, $r_A=|r_A|\exp(i \phi_A)$,  $\alpha = f_p \in \mathrm{Re}$,
and $\beta = f_s \exp(i \psi)$. Note that the expressions in
Eqs.~(\ref{TSR}) and (\ref{TLSR}) have a similar form as those in a
semi-infinite medium~\cite{Aiello2008}, although the refection
coefficients are significantly different.

We now discuss the transverse shifts in the tunneling field. After
substituting Eq.~(\ref{fieldt}) into Eq.~(\ref{centroid}), we obtain
\begin{eqnarray}
\Delta y_t &=&-\frac{1}{k_0}\frac{f_p f_s \cot\theta_i }{|t_p|^2
f_p^2
+|t_s|^2 f_s^2}\bigl[\bigl(|t_p|^2 + |t_s|^2 \bigr)\sin\psi\nonumber\\
&&-2\eta|t_p||t_s|\sin(\psi - \varphi_p + \varphi_s )
\bigr]\label{TST},
\end{eqnarray}
\begin{equation}
\delta y_t=\frac{z_t}{k_0 z_{Ry}}\frac{f_p
f_s(|t_p|^2-|t_s|^2)\cot\theta_i\cos\psi}{|t_p|^2 f_p^2+|t_s|^2
f_s^2}\label{TLST},
\end{equation}
where $t_A=|t_A|\exp(i \varphi_A)$.  From Eqs.~(\ref{TLSI}),
(\ref{TLSR}), and (\ref{TLST}), we find that the $z_a$-dependent
terms can be regarded as a small angle shift inclining from the axis
of beam centroid:
\begin{equation}
\Delta\theta_{ay}=\delta y_a/z_a\label{RTAS}.
\end{equation}
This angle divergences means that the Snell's law cannot accurately
describe the beam refraction phenomenon~\cite{Duval2006}. It should
be mentioned that $\delta y_{r}$ and $\delta y_{t}$ are given by
functions of Rayleigh lengths and $\cos\psi$. Hence, the transverse
angular shifts can be regarded as the combined contribution of
diffraction and polarization~\cite{Luo2009}. In general, a linearly
polarized Gaussian beam can be regarded as a superposition of two
circularly polarized components. As a result, an initially linearly
polarized beam splits into two circularly polarized components in
opposite directions due to the SHE of light~\cite{Bliokh2006a}.

\begin{figure}
\includegraphics[width=8cm]{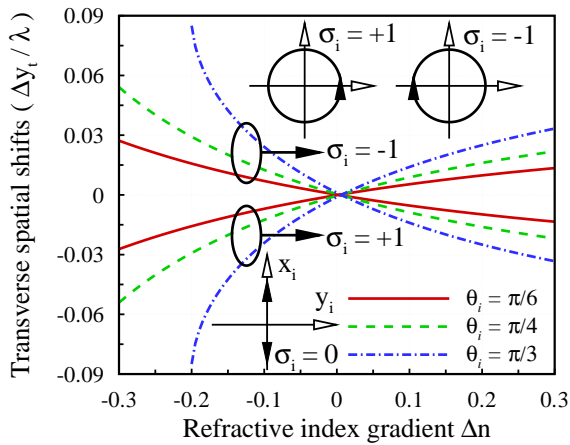}
\caption{\label{Fig2} (color online) The normalized transverse
spatial shifts $\Delta{y_t}/\lambda$ versus refractive index
gradient $\Delta{n}$. The incident angles of the beam are choose as
$\theta_i=\pi/6$ (solid lines), $\theta_i=\pi/4$ (dashed lines), and
$\theta_i=\pi/3$ (dashed-dotted lines). The gap thickness is chosen
as $d=0.2\lambda$. The insets show that the linear polarization $\sigma_i=0$
can be regarded as a superposition of two circular polarization
components $\sigma_i=+1$ and $\sigma_i=-1$.}
\end{figure}

In the SHE of light, the refractive index gradient plays the role of
the electric potential gradient. Hence, we attempt to examine what
roles the refractive index gradient of the two prisms (i.e.,
$\Delta{n}=n_2-n_1$) plays in the photon tunneling. Without loss of
generality, we assume $n_1=1.5$ and the corresponding critic angle
is $\theta_{ic}=41.8^{\circ}$. Figure \ref{Fig2} shows the
normalized transverse spatial shifts for various refractive index
gradients. We first consider the photon tunnels through the air gap
from a high-refractive-index prism to a low-refractive-index prism
($\Delta{n}<0$). For the left circularly polarized component
$\sigma_i=+1$, the transmitted field exhibits a negative transverse
shift. For the right circularly polarized component $\sigma_i=-1$,
also presents a transverse shift, but in an opposite direction. We
next consider the photon tunneling from a low-refractive-index prism
to a high-refractive-index prism ($\Delta{n}>0$). We find that the
left circularly polarized component presents a positive shift, while
the right circularly polarized component exhibits a negative shift.
For the left or the right circularly polarized component, the SHE of
light is reversed when the refractive index gradient is inverted.

\begin{figure}
\includegraphics[width=8cm]{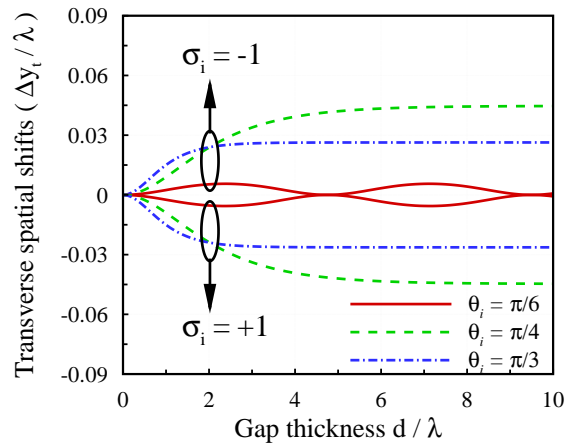}
\caption{\label{Fig3} (color online) The normalized transverse
spatial shifts $\Delta{y_t}/\lambda$ versus normalized gap thickness
$d/\lambda$. The incident angles of the beam are choose as
$\theta_i=\pi/6$ (solid lines), $\theta_i=\pi/4$ (dashed lines), and
$\theta_i=\pi/3$ (dashed-dotted lines). The refractive index
gradient is chosen as the symmetric case $\Delta{n}=0$.}
\end{figure}

In conventional photon tunneling
experiments~\cite{Steinberg1993,Balcou1997}, the barrier is
constructed by a symmetric structure, i.e., $\Delta{n}=0$. It would
be interesting to give more details about the SHE of light in this
standard case. Figure~\ref{Fig3} plots the transverse spatial shifts
as a function of the gap thickness. The transverse shifts present
with the increase of the gap thickness. This interesting phenomenon
is due to optical resonance effect in the air barrier. Therefore,
such a mechanism provides a possible way to amplify the SHE of light.
In general, the gap thickness in tunneling experiment is a small value, which
leads to the polarization-dependent transverse shifts are a tiny
effect. This is a possible reason why the SHE of light escapes
measurement in photon tunneling experiments. Note that the SHE of
light can be evidently enhanced by increasing the refractive index
gradient between the two prisms (Fig.~\ref{Fig2}). The recent advent of metamaterial
whose refractive index can be tailored
arbitrarily~\cite{Smith2004,Pendry2006,Shalaev2007} seems to be a
good candidate. A further point should be mentioned: The transverse
spatial shifts in the conventional SHE experiment are just a few tens of
nanometers~\cite{Hosten2008}. Fortunately, the transverse spatial shifts in
the tunneling case (dashed and dashed-dotted lines) are much larger
than those in the transmission case (solid lines). Thus, the
tunneling mechanism provides us an alternative way to enhance the
SHE of light.

\begin{figure}
\includegraphics[width=8cm]{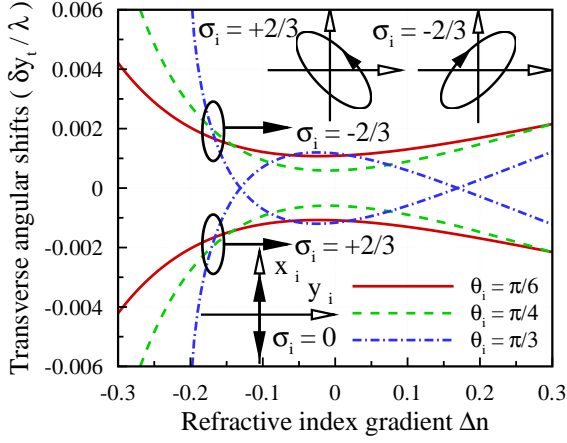}
\caption{\label{Fig4} (color online) The normalized transverse
angular shifts $\delta{y_t}/\lambda$ versus refractive index
gradient $\Delta{n}$. Polarization parameters are $\sigma_i=+2/3$
and $\sigma_i=-2/3$. The angular shifts are plotted in the plane
$z_t=2z_R$. Other parameters are chosen to be the same as in
Fig.~\ref{Fig2}. The insets show that the linearly polarized beam
with $\sigma_i=0$ can be regarded as a superposition of two
elliptically polarization components with $\sigma_i=+2/3$ and
$\sigma_i=-2/3$.}
\end{figure}

We next explore the polarization-dependent transverse angular
shifts. From Eq.~(\ref{TLST}) we find that only the special linearly
or the elliptically polarized beam exhibits a transverse angular
shift. In this case, a linearly polarized Gaussian beam can be
regarded as a superposition of two elliptically polarized components
as shown in Fig.~\ref{Fig4}. For a left-elliptical component
$\sigma_i=+2/3$ and a right-elliptical component $\sigma_i=-2/3$,
the centroid of tunneling field presents an opposite angular shifts.
For the left- or right-elliptical component, whether the transverse
angular shift is positive or negative depends on the incident angle
and the refractive index gradient. It should be noted that the
transverse angular shift is significantly different from the
longitudinal one~\cite{Balcou1997,Merano2009}, which is
polarization-dependent.

\begin{figure}
\includegraphics[width=8cm]{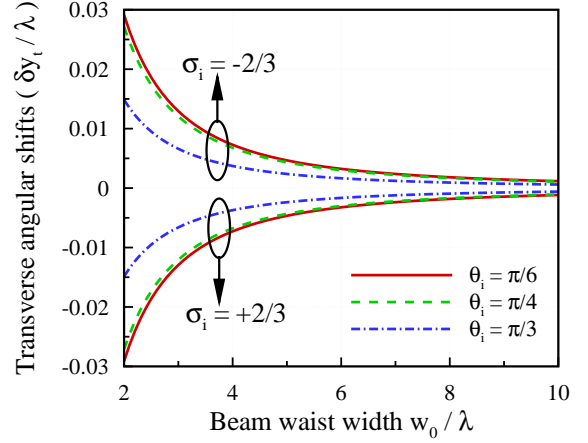}
\caption{\label{Fig5} (color online) The normalized transverse
angular shifts $\delta{y_t}/\lambda$ versus normalized  beam waist
width $w_0/\lambda$. Polarization parameters are $\sigma_i=+2/3$ and
$\sigma_i=-2/3$. The refractive index
gradient is chosen as the symmetric case $\Delta{n}=0$.
Other parameters are chosen to be the same as in
Fig.~\ref{Fig4}.}
\end{figure}

It is known that the transverse angular shifts can be regarded as the combined
contribution of diffraction and polarization. Therefore, it is
necessary to examine the role of the beam waist in the SHE of light.
The normalized transverse angular shifts versus the beam waist width are
plotted in Fig.~\ref{Fig5}. It is clearly shown that the transverse
angular shift increases with the decrease of the beam waist width.
For a small beam waist, the transverse angular shift can be
noticeably enhanced in both transmission and tunneling situations.
This would be of interest for the application to nano-photonics. It
should be pointed out that the paraxial propagation model only deals with beams
whose transverse dimension is much larger than a
wavelength~\cite{Lax1975}. When the beam waist is small compared
with the wavelength, a nonparaxial propagation model should be
developed.

For the purpose of revealing the physics underlying the SHE, we
focus on the momentum conservation laws which govern both the
spatial and the angular shifts. The monochromatic beam can be
formulated as a localized wave packet whose spectrum is arbitrarily
narrow~\cite{Bliokh2007}. Let $a$th wave packet include $N_a$
photons, i.e., its field energy is $W_a=N_a \omega$. The linear
momentum of the $a$th wave packet is $\mathbf{p}_a=N_a\mathbf{k}_a$
(we set $\hbar=c=1$), and the linear momentum conservation law for
$y$ components is $p_{iy}=p_{ry}+p_{ty}$, where $p_{iy}=0$,
$p_{ry}=N_r k_r \Delta\theta_{ry}$, and $p_{ty}=N_t k_t
\Delta\theta_{ty}$. In the tunneling process, the total number of
photons remains unchanged: $N_r+N_t=N_i$. As a result, the
transverse angular shifts fulfill the conservation law for $y$
components of the linear momentum:
\begin{equation}
-n_1 Q_r \Delta\theta_{ry}+ n_2 Q_t\Delta\theta_{ty}=0\label{PY}.
\end{equation}
Here, $Q_r=N_r/N_i$ and $Q_t=N_t/N_i$ are energy reflection and
energy transmission coefficients, respectively. In the frame of
classic electrodynamics, the corresponding coefficients can be
written as
\begin{equation}
Q_r=f_p^2|r_p|^2+f_s^2|r_s|^2\label{Qr},
\end{equation}
\begin{equation}
Q_t=n_2\eta(f_p^2 |t_p|^2+f_s^2|t_s|^2)/n_1\label{Qt}.
\end{equation}
To verify the conservation law, we need to determine the transverse
angle divergences, which can be obtained from Eq.~(\ref{RTAS}) as
\begin{equation}
\Delta\theta_{ry}=\frac{1}{k_0 z_R}\frac{f_p f_s  (|r_p|^2 -
|r_s|^2)\cot\theta_i \cos\psi}{|r_p|^2 f_p^2+|r_s|^2
f_s^2}\label{ATLSR},
\end{equation}
\begin{equation}
\Delta\theta_{ty}=\frac{1}{k_0 z_{Ry}}\frac{f_p
f_s(|t_p|^2-|t_s|^2)\cot\theta_i\cos\psi}{|t_p|^2 f_p^2+|t_s|^2
f_s^2}\label{ATLST}.
\end{equation}
After substituting Eqs.~(\ref{Qr})-(\ref{ATLST}) into
Eq.~(\ref{PY}), we find that the conservation law still holds. Thus,
the transverse angular shifts are governed by the total linear
momentum conservation law.

We proceed to explore the total angular momentum conservation law.
The $z$ component of total angular momentum $L_{az}$ for $a$th wave
packet can be represented as a sum of the extrinsic orbital angular
momentum $J_{az}$ and intrinsic spin angular momentum $S_{az}$,
i.e.,  $L_{az}=J_{az}+S_{az}$. The $z$ component of the orbital
angular momenta are given by $J_{iz}=0$, $J_{rz}=-\Delta y_r k_r
\sin\theta_r$, and $J_{tz}=-\Delta y_t k_t \sin\theta_t$. The $z$
components of the spin angular momenta are given by
$S_{iz}=\sigma_i\cos\theta_i$, $S_{rz}=\sigma_r \cos\theta_r$, and
$S_{tz}=\sigma_t \cos\theta_t$ for incident, reflected, and
tunneling wave packets, respectively. The total angular momentum can
be written as
\begin{equation}
L_{iz}=\sigma_i \cos\theta_i,\label{LIZ}
\end{equation}
\begin{equation}
L_{rz}=-\Delta y_r k_r \sin\theta_r+\sigma_r
\cos\theta_r\label{LRZ},
\end{equation}
\begin{equation}
L_{tz}=-\Delta y_t k_t \sin\theta_t+\sigma_t
\cos\theta_t\label{LTZ}.
\end{equation}
Here, the polarization degrees for the $a$th wave packet is
respectively described by
\begin{equation}
\sigma_i=2 f_p f_s \sin\psi,
\end{equation}
\begin{equation}
\sigma_r=\frac{2 f_p f_s|r_p||r_s|\sin[\psi - (\phi_p - \phi_s
)]}{|r_p|^2  f_p^2 +|r_s|^2 f_s^2},
\end{equation}
\begin{equation}
\sigma_t=\frac{2 f_p f_s|t_p||t_s|\sin[\psi - (\varphi_p - \varphi_s
)]}{|t_p|^2  f_p^2 +|t_s|^2 f_s^2}.
\end{equation}
After substituting Eqs.~(\ref{TSR}) and (\ref{TST}) into
Eqs.~(\ref{LRZ}) and (\ref{LTZ}), we find that the transverse  spatial shifts
fulfill the conservation law for total angular momentum:
\begin{equation}
Q_r L_{rz}+ Q_t L_{tz}=L_{iz}.\label{totalam}
\end{equation}
The transverse shifts governed by the total momentum conservation
law, provide us a new way to clarify the nature of photon tunneling
effect in three dimensions. Note that the magnitude of the SHE of
light can be substantially enhanced by involving  higher-order
angular momentum
beams~\cite{Allen1992,Bliokh2006b,Bliokh2009a,Fadeyeva2009,Merano2010}. Future
research is needed to investigate the orbital Hall effect of
light in photon tunneling regime.

To obtain a clear physical picture of the SHE of light in photon
tunneling, we attempt to perform analyses on the $z$ component of
the total angular momentum for each of individual photons, i.e.,
$l_{iz}=l_{tz}$~\cite{Onoda2006}. The total angular momentum law for
single photon is given by
\begin{equation}
-\Delta y_t k_t \sin\theta_t+\sigma_t \cos\theta_t=\sigma_i
\cos\theta_i.
\end{equation}
When the photons tunnel through the air-gap barrier from a
low-refractive-index prism to a high-refractive-index prism
($n_1<n_2$), the incident angle is larger than the transmitted angle
$\theta_i>\theta_t$. The linearly polarized wave packet can be
represented as a superposition of equal $\sigma_i=+1$ and
$\sigma_i=-1$ photons. For the $\sigma_i=+1$ photons, the $z_t$
component of spin angular momentum $\sigma_t \cos\theta_t$ increases
after entering the second prism. Because of the conservation law,
the total angular momentum must remain unchanged. To conserve the
total angular momentum, the photon must move to the direction $+y$
($\Delta y_t>0$). For the $\sigma_i=-1$ photons, the $z$ component
of spin angular momentum $\sigma_t \cos\theta_t$ decreases. In this
case, the photons must move to the direction $-y$ ($\Delta y_t<0$).
When the photons tunnel through the air-gap barrier from a
high-refractive-index prism to a low-refractive-index prism
($n_1>n_2$), the incident angle is less than the transmitted angle
$\theta_i<\theta_t$. As expected, the transverse spatial shifts
exhibit a reversed version. This gives a very simple way to
understand how light exhibits SHE in the photon tunneling.

Chiao and Steinberg~\cite{Chiao1991,Steinberg1994} have studied
systematically the equivalence between the FTIR equation and the
Schr\"{o}dinger one. Theoretical studies have shown indeed that the
physics of tunneling is essentially identical for classical light
waves and quantum mechanical wave functions.  The SHE of light has
been verified experimentally in beam refraction~\cite{Hosten2008}
and in glass cylinder with refractive index
gradient~\cite{Bliokh2008}. However, direct observation of the SHE
is still remain an open challenge in condensed matter
physics~\cite{Murakami2003,Sinova2004,Wunderlich2005} and
high-energy physics~\cite{Berard2006,Gosselin2007}. This opened up
the opportunity to perform the SHE in tunneling experiments, easier
to perform and interpret than those with electron waves. Because of
the close similarity in condensed matter, high-energy physics,
plasmonics~\cite{Gorodetski2008,Gorodetski2009,Vuong2010,Herrera2010},
and optics, the SHE of light in photon tunneling will provide
indirect evidence for other physical systems.

\section{Conclusions}
In conclusion, we have resolved the breakdown of angular momentum
conservation on two-dimensional photon tunneling by considering SHE
of light. This interesting effect manifests itself as
polarization-dependent transverse shifts when the wave packet
tunnels through a prism-air-prism barrier. For the left or the right
circularly polarized component, the transverse shift can be
modulated by altering the refractive index gradient of the two
prisms. We have demonstrated that the SHE in the conventional beam
refraction can be evidently enhanced via photon tunneling. The
physics underlying the SHE of light is that the transverse angular
shifts satisfy the total linear momentum conservation law, and the
transverse spatial shifts fulfill the total angular momentum
conservation law. The study of the SHE of light would make a useful
contribution to clarify the nature of photon tunneling in three
dimensions. It is well known that the scanning tunneling microscope
is a powerful instrument for imaging surfaces at the atomic
level~\cite{Novotny2006}. In general, introducing the SHE of light
into the scanning tunneling microscope will open the possibility for
developing new nano-photonic devices.

\begin{acknowledgements}
We are sincerely grateful to the anonymous referee, whose comments
have led to a significant improvement of our paper. This research
was supported by the National Natural Science Foundation of China
(Grants Nos. 10804029, 10974049, and 11074068).
\end{acknowledgements}

\end{document}